\documentclass[letterpaper,twocolumn,10pt]{article}
\usepackage{usenix2024_SOUPS}

\usepackage{tikz}
\usepackage{amsmath}
\usepackage{xurl}

\usepackage{filecontents}
\usepackage{booktabs}
\begin{document}

\date{}

\title{\Large \bf From Preventive to Reactive: How AI Coding Assistants Transform Developers' Security Awareness}

\def\plainauthor{Bappy et al.}


\author{
{\rm Faisal Haque Bappy}$^{\star}$,
{\rm Tahrim Hossain}$^{\star}$,
{\rm Sidratul Muntaher Meheraj}$^{\circ}$,
{\rm Annoor Sharara Akhand}$^{\circ}$,\\
{\rm Tasfia Tabassum}$^{\circ}$,
{\rm Tarannum Shaila Zaman}$^{\star}$,
{\rm Raiful Hasan}$^{\diamond}$,
{\rm Tariqul Islam}$^{\star}$\\[0.5em]
$^{\star}$University of Maryland Baltimore County, 
$^{\circ}$University of Dhaka, 
$^{\diamond}$Kent State University
}

\maketitle
\thecopyright

\begin{abstract}
AI coding assistants are now central to professional software development, yet their impact on how developers think about and practice security remains poorly understood. While prior work has documented vulnerability rates in AI-generated code, a more fundamental question persists: how do these tools transform security awareness in authentic, ongoing development practice?
We conducted semi-structured interviews with 15 professional software engineers and observed them completing security-relevant coding tasks with AI assistance, spanning 3 experience cohorts defined by their relationship to AI tools during professional formation.
We find that AI coding assistants reorganize rather than eliminate security thinking, shifting it from the act of writing code to the act of reviewing it. This transition from preventive to reactive security is structurally encouraged by interaction models that frame code generation as a functional task, leaving security as an afterthought. Notably, none of our coding session participants specified security requirements in their initial prompts, even when they possessed the relevant knowledge, revealing a decoupling of security awareness from security behavior. We further document informal coping strategies developers had independently invented to manage AI security risk, none of which are supported by current tools or organizations, and find that the experience cohort did not reliably predict security performance. This paper contributes a practice-grounded account of how AI-assisted development reshapes the human side of secure coding, offering empirical foundations for the design of more security-aware tools, training programs, and organizational policies.
\end{abstract}

\section{Introduction}

Software development is changing faster than our understanding of its consequences. AI coding assistants, including GitHub Copilot~\cite{github_copilot}, ChatGPT~\cite{openai_chatgpt}, and Claude~\cite{anthropic_claude}, have become part of the daily infrastructure of professional engineering work within just a few years \cite{kalliamvakou2022writing, peng2023impact, chen2021evaluating}. Developers now routinely delegate code generation to these tools, accepting suggestions, iterating on prompts, and shipping the results at a pace that was previously impossible \cite{barke2023grounded, vaithilingam2022expectation}. The productivity case for this shift is well established \cite{peng2023impact}. The security case is considerably less so.

What we already know is concerning. Pearce et al. \cite{10.1145/pearce2022asleep} found that roughly 40\% of code generated by GitHub Copilot across high-risk vulnerability scenarios was exploitable, spanning weakness categories well documented in the OWASP Top Ten \cite{owasp2021}. Perry et al. \cite{perry2023users} found that developers with access to an AI assistant produced less secure code than those without, and were simultaneously more likely to believe their code was secure. This combination of increased vulnerability and increased confidence points to something more troubling than simple tool failure: AI coding assistants may be quietly reshaping how developers perceive and engage with security, not just in the code they produce, but in the habits of mind they bring to reviewing it.

Yet the existing literature stops short of explaining how or why this happens in practice. Technical evaluations measure code outputs in controlled settings \cite{10.1145/pearce2022asleep, perry2023users}, but say little about the cognitive and behavioral processes that produce those outputs in authentic, ongoing development work. The AI-assisted programming literature has documented interaction patterns \cite{barke2023grounded}, usability expectations \cite{vaithilingam2022expectation}, and productivity outcomes \cite{peng2023impact, xu2022systematic}, but has largely treated security as peripheral. The developer security literature, meanwhile, has shown that security is difficult to sustain even without AI in the picture: developers avoid static analysis tools because the warnings lack actionable context \cite{johnson2013don}, the information sources they consult shape the vulnerabilities they introduce \cite{acar2016you}, and the shift away from manual information foraging \cite{brandt2009two} toward automated suggestion carries risks the field has not yet fully worked through. The automation trust literature adds another layer of concern, showing that people exhibit systematic biases toward trusting automated outputs \cite{parasuraman1997humans, lee2004trust}, biases that tend to grow stronger as tools perform well on average, making occasional failures harder to notice and more consequential when they slip through \cite{kocielnik2019will, bansal2019updates}.

None of these bodies of work, individually or together, provides a qualitative account of how AI coding assistants transform security awareness as a lived practice. The central question this paper addresses is: \textit{How do AI coding assistants influence the ways professional software engineers approach security, and how do these effects differ based on developers’ levels of exposure to AI tools during their professional formation?}

To answer this, we conducted a mixed-methods qualitative study with 15 professional software engineers, combining semi-structured interviews with observed coding sessions in which participants completed security-relevant implementation tasks using AI assistants. Participants were drawn from 3 experience cohorts: Pre-AI developers who built their foundational practices before AI tools were widely available, AI-era developers who adopted AI mid-career, and AI-native developers who have used AI throughout their professional lives. Our analysis follows established thematic analysis procedures \cite{braun2006using, charmaz2014constructing}, combining iterative coding of interview transcripts with systematic behavioral observation of the coding sessions.

Our findings highlight several interconnected patterns that together point to a meaningful shift in how security awareness operates when AI enters the development workflow:

\begin{itemize}

    \item \textbf{Security thinking moves from authorship to review.} Security considerations, traditionally exercised during the act of writing code \cite{mcgraw2012software}, now appear to be displaced downstream into the review phase. As a result, they are introduced later in the development cycle, where they must compete with time pressure and are applied less consistently. Notably, none of the participants in our coding sessions included security requirements in their initial prompts, even those who had just articulated the relevant vulnerability concerns in the interview, suggesting that security knowledge and security prompting behavior can come apart even among aware engineers.

    \item \textbf{Security outcomes depend heavily on what engineers bring to the interaction.} AI coding assistants do not apply security constraints by default and produce functionally plausible but security-incomplete code unless explicitly directed otherwise. This places the burden of security squarely on the individual engineer's knowledge and prompting habits, creating uneven outcomes that are difficult to address through tool adoption alone.

    \item \textbf{Trust is poorly calibrated to security sensitivity.} Consistent with the broader automation literature \cite{parasuraman1997humans, lee2004trust}, participants applied similar levels of trust to both boilerplate (i.e., standardized template or scaffolding code reused across projects) and security-critical code, even though the tool provided no indication that the risks associated with these categories were different. Importantly, experience cohort did not reliably predict security performance during the coding sessions, suggesting that years of experience is a limited proxy for security-aware AI use.

    \item \textbf{Developers have built their own coping strategies that tools and organizations have not yet caught up with.} Participants described independently developing practices such as restricting agent permissions, constructing security-oriented prompt templates, and explicitly asking AI whether any security concerns remained after completing a functional implementation. These strategies were effective when applied, but they existed entirely outside any formal guidance or tool support, leaving their consistent use to individual initiative.

\end{itemize}

What these findings share is a common structural explanation: the shift from preventive to reactive security awareness is not primarily a knowledge problem or a carelessness problem. It is a structural outcome of how AI coding assistants are currently designed and adopted \cite{anderson2010security, ortloff2025replication}, and addressing it will require intervention at the level of tools, training, and organizational practice. This paper takes a step toward that intervention by documenting, in developers' own words and actions, how the shift is already underway.

\section{Related Work}

Our study draws on and critically engages with four bodies of literature: AI-assisted software development, developer security behavior and awareness, trust and reliance in AI systems, and the intersection of AI tools with security practices. While substantial work exists in each area, we identify significant gaps that our qualitative investigation addresses.

\subsection{AI-Assisted Software Development}

The emergence of large language models trained on code repositories has fundamentally altered how software is written \cite{chen2021evaluating, xu2022systematic}. Tools like GitHub Copilot, ChatGPT, and Claude Code now offer real-time code suggestions within development environments, and empirical evaluations have demonstrated measurable productivity gains. Peng et al. \cite{peng2023impact} found that developers completed tasks significantly faster with Copilot, while Vaithilingam et al. \cite{vaithilingam2022expectation} documented mixed results depending on task complexity and developer experience.

However, the existing literature on AI coding assistants suffers from a narrow fixation on productivity metrics. Weisz et al. \cite{weisz2025examining} investigated IBM's AI coding assistant and found that developers mainly used it for code comprehension and explanation rather than generation, challenging the dominant framing of these tools as mere code producers. Critically, their participants described coding with AI as a collaborative process involving shared authorship and responsibility, yet the study stopped short of examining what happens to security-specific responsibilities in this shared model. Pu et al. \cite{pu2025assistance} introduced Codellaborator, a proactive AI programming assistant, and acknowledged the tension between AI proactivity and user control, finding that users feared losing ownership and understanding. Their recommendation that AI tools should slow down generation to help users grasp code logic is well taken, but the study did not investigate whether this loss of understanding extends to security-critical reasoning, a gap our work directly addresses.

The growing literature on proactive AI assistants further illustrates this blind spot. Chen et al. \cite{chen2025need} explored proactive chat-based AI assistants for programming and confirmed productivity improvements, but their design considerations around user control and context awareness entirely omit security as a dimension. Similarly, ChainForge \cite{arawjo2024chainforge} and ChainBuddy \cite{zhang2025chainbuddy} advanced prompt engineering toolkits for LLM pipelines, yet neither considered how developers verify the security properties of AI-generated outputs. Zhang et al. \cite{zhang2025chainbuddy} even noted that participants using ChainBuddy became dependent on the AI and often missed AI-generated errors, a finding with alarming implications for security that the authors did not explore. Kretzer et al. \cite{kretzer2025closing} demonstrated LLM-powered GUI prototyping that reduced manual effort, but the question of whether reduced manual engagement also reduces security scrutiny remained unasked.

Chen et al. \cite{chen2024learning} studied how novices and experts learn agent-based modeling with LLM companions and found that novices need substantially more support in planning, understanding, and debugging. Their finding that LLMs should adapt guidance based on skill level aligns with our observation that developers across experience cohorts interact with AI coding assistants in different ways. However, the prior study did not treat security knowledge as a distinct dimension of expertise that warrants adaptation. Across this body of work, the consistent omission of security as a variable of interest is itself a finding: the field has implicitly treated security as orthogonal to AI-assisted development, an assumption our study challenges.

\subsection{Developer Security Behavior and Awareness}

Security has long been recognized as a persistently challenging dimension of software development \cite{mcgraw2012software, anderson2010security}. Despite decades of research and extensive documentation of secure alternatives, common vulnerability categories such as injection flaws, broken authentication, and sensitive data exposure remain stubbornly prevalent \cite{owasp2021}. Johnson et al. \cite{johnson2013don} found that static analysis tools fail because developers misinterpret warnings, lack context for severity assessment, or face integration friction. Acar et al. \cite{acar2016you} showed that the choice of information source significantly affects the security of code developers produce, with Stack Overflow's copy-paste culture contributing to vulnerability propagation.

The replication study by Ortloff et al. \cite{ortloff2025replication} on expert versus non-expert security practices provides important longitudinal context: over the past decade, non-experts have increasingly adopted expert-level practices like two-factor authentication, VPNs, and ad blockers, driven largely by improved usability and institutional mandates. However, the same study identified a troubling consequence of automation: as security processes become less visible, users lose their sense of control and awareness. This tension between automation benefit and awareness erosion is precisely what we observe in the AI coding assistant context, but manifested at a more granular, code-level scale.

Mowar et al. \cite{mowar2025codea11y} developed CodeA11y, a Copilot extension that reinforced accessible coding practices, demonstrating that AI-generated code still requires human review and that developers, especially novices, miss important issues without explicit guidance. While their focus was accessibility rather than security, the underlying dynamic is transferable: when AI handles generation, developers' capacity to identify non-functional quality attributes degrades unless the tool actively surfaces them.

Widder et al. \cite{widder2023power} examined the ethical concerns of software engineers more broadly, finding that engineers' power to resolve concerns was constrained by financial precarity, workplace culture, and organizational incentives. Their finding that attempts to address ethical issues were “often psychologically taxing and ineffective” helps explain why developers may deprioritize the security review of AI generated code. Such review introduces additional friction into workflows that AI tools were adopted specifically to streamline.

\subsection{Trust, Reliance, and Transparency in AI Systems}

Understanding how developers calibrate their trust in AI coding assistants requires engaging with the broader trust and automation literature. Parasuraman and Riley's foundational work on automation \cite{parasuraman1997humans} identified systematic biases toward trusting automated outputs, while Lee and See \cite{lee2004trust} established frameworks for appropriate reliance. In AI-specific contexts, the question of when trust is justified has received renewed attention.

Manzini et al. \cite{manzini2024should} argued that user trust in advanced AI assistants is only justified when there is evidence of both the AI's competence and its alignment with user values, requiring interventions at the assistant design, organizational, and governance levels. This theoretical framework exposes a critical gap in current AI coding assistant deployments: developers have limited evidence of either competence or alignment when it comes to security, yet trust levels remain high. Wang et al. \cite{wang2024investigating} investigated trust formation in AI code generation tools specifically, finding that developers base trust on perceived ability, integrity, and benevolence, and that trust varies with task complexity and stakes. Their finding that current tools lack affordances for validating output quality directly supports our observation that developers lack mechanisms for security-specific trust calibration.

Pareek et al. \cite{pareek2024trust} showed that explicit trust repair strategies play a significant role in restoring trust after AI failures. Among the strategies they evaluated, model updates were the most effective in rebuilding trust, while denial tended to deepen distrust. For security contexts, this has practical implications: when AI-generated code introduces a vulnerability, the absence of any acknowledgment or repair mechanism may lead to sustained inappropriate trust. Lee et al. \cite{lee2024one} found that presenting users with multiple, potentially inconsistent LLM outputs reduced overreliance by making the model’s limitations more visible. In fact, exposure to inconsistencies improved user comprehension rather than undermining it. This suggests that the current design pattern of presenting a single ``best'' code suggestion may actively suppress the critical evaluation that security review demands.

Serafini et al. \cite{serafini2025exploring} conducted the most directly relevant experimental work, testing how security prompts and AI warning messages affect developers' behavior when using insecure AI-generated code. Their findings are sobering: developers frequently trusted insecure AI suggestions like MD5 hashing and failed to verify outputs properly. While security prompts and guidelines reduced the use of insecure code, the study was conducted in a controlled experimental setting with manipulated ChatGPT responses. Our qualitative investigation complements this by examining how trust and verification behaviors manifest in developers' authentic, ongoing practices rather than one-time experimental tasks.

The broader literature on AI transparency further contextualizes our findings. Chan et al. \cite{chan2024visibility} argued that autonomous AI agents introduce unique risks requiring visibility measures, including agent identifiers, real-time monitoring, and activity logging. Turri et al. \cite{turri2024transparency} demonstrated through a case study that current methods for identifying user transparency needs are too narrow, requiring multi-method approaches to capture diverse stakeholder needs. For AI coding assistants, neither visibility into agent decision-making nor transparency about security-relevant choices is currently provided, leaving developers to rely on heuristic judgment.

\subsection{AI at the Intersection of Security}

A small but growing body of work examines AI tools within security-specific contexts, though this literature remains fragmented. The SmishX system \cite{wang2025can} achieved high accuracy in SMS phishing detection with explainable AI, but identified the risk of overreliance on AI and difficulty in error correction when the AI misclassifies messages. These challenges mirror what we observe in code generation: high baseline performance breeds complacency, and the cognitive cost of overriding AI suggestions discourages critical evaluation.

Research on integrating LLMs into security incident response \cite{kramer2025integrating} found that experts preferred AI-assisted summaries in the majority of cases, but over half of the summaries required factual corrections. The gap between perceived utility and actual reliability is a recurring theme in AI assisted security work. Our study shows that this gap also appears in code generation settings, where developers rely on AI outputs for security sensitive tasks without sufficient verification.

Studies on AI adoption in cybersecurity settings reveal persistent barriers. Research examining security experts' perspectives on augmented intelligence \cite{roch2024navigating} found that only about a third of organizations use or plan to use AI for protection, with experts requiring adaptability, transparency, determinism, and openness as prerequisites for trusted collaboration. Research on AI adoption in industrial control systems \cite{fung2025adopting} similarly found that human tasks relying on experience and intuition are better suited for AI-assisted rather than AI-automated support. These findings from security operations contexts reinforce our argument that security critical code generation requires human in the loop oversight. However, current AI coding assistants do not provide mechanisms to distinguish between security sensitive and routine code generation tasks.

Madaio et al. \cite{madaio2024learning} investigated how AI practitioners learn about responsible AI, finding that existing materials are predominantly oriented toward technical and checklist-based approaches, while practitioners aspire to sociotechnical understanding. This disconnect between available learning resources and needed competencies parallels what we observe: developers need to understand the security implications of AI-assisted development holistically, but available guidance remains narrowly technical.

\subsection{Gaps Addressed by This Work}
The literature reveals three critical gaps. First, research on AI coding assistants has overwhelmingly prioritized productivity, usability, and learning outcomes while treating security as peripheral. Second, the developer security literature has not yet reckoned with how AI-mediated code generation transforms the cognitive and practical dimensions of security awareness. Third, while trust and transparency research provides useful frameworks, it has not been applied to understanding how developers calibrate security-specific trust in AI-generated code across different experience levels.

Our study bridges these gaps through a qualitative investigation that foregrounds security awareness as the central variable of interest, examines how it transforms rather than simply persists or disappears across developer experience cohorts, and grounds its findings in developers' authentic practices and reasoning rather than controlled experimental tasks.

\section{Methodology}

The study consisted of 2 components conducted in sequence during a single session lasting approximately 45 to 60 minutes. In the first component, we conducted a conversational semi-structured interview exploring participants' experiences maintaining security in their day-to-day development work, their familiarity with AI coding assistants, and how (if at all) these tools had changed how they think about security. In the second component, participants completed a Think-Aloud coding task using Gemini CLI~\cite{google2025geminicli} with the Gemini 2.5 Pro model, during which we observed their prompting strategies, decision-making processes, and any security-relevant behaviors or omissions. Both components were designed to surface not just what developers \textit{say} about security, but what they \textit{do} when AI assistance is readily available.

\subsection{Participants}

We recruited 15 professional software developers between November 2025 and January 2026 using snowball~\cite{goodman1961snowball} and purposive sampling~\cite{patton1990qualitative} o ensure diversity in roles, experience levels, and familiarity with AI coding tools. All participants were industry professionals; no students or university employees were included. Participation was voluntary and uncompensated. We did not collect employer or organizational identifiers, as discussing AI usage and security practices may carry workplace sensitivity. Because our study examines how developers engage with AI-assisted code generation, we grouped participants according to when they entered the profession relative to the emergence of large language model tools, allowing us to analyze how professional formation shapes interaction with AI systems. Participants were categorized into 3 AI relationship cohorts: \textit{Pre AI} developers began their professional software engineering careers before LLM based tools became widely available; \textit{AI era} developers were trained prior to this shift but entered the workforce as AI tools became mainstream; and \textit{AI native} developers both trained and began their careers entirely within the era of accessible AI coding assistants. Cohort boundaries were set at pre-2022, 2022 to 2023, and 2024 onward, corresponding to the period before, during, and after widespread LLM tool adoption.

Participants ranged from entry-level to mid-level engineers across roles, including full-stack development, machine learning engineering, frontend engineering, research and development, and general software engineering. Table~\ref{tab:participants} provides a complete summary of participant demographics and study characteristics. Thematic saturation was reached at approximately 13 participants, after which the final 2 interviews confirmed existing themes without introducing new ones.

\begin{table*}[htbp]
\centering
\caption{Participant Demographics and Study Characteristics}
\label{tab:participants}
\resizebox{0.96\textwidth}{!}{%
\begin{tabular}{lllllll}
\hline
\textbf{ID}  & \textbf{Gender\textsuperscript{*}} & \textbf{Role}                                      & \textbf{Experience} & \textbf{AI Cohort}   & \textbf{Coding Task Selected}             \\ \hline
P1  & M      & Mid-Level Full-Stack Developer             & 5 yrs      & Pre-AI      & Feature Implementation    \\
P2  & F      & Full-Stack Engineer                        & 2 yrs      & Pre-AI      & Project Planning          \\
P3  & M      & ML Engineer L2                             & 4 yrs      & Pre-AI      & Project Planning          \\
P4  & M      & Mid-Level R\&D Engineer                   & 2 yrs      & Pre-AI      & Debugging                 \\
P5  & M      & Mid-Level Full-Stack Developer             & 4 yrs      & Pre-AI      & All Three Tasks           \\
P6  & M      & Mid-Level Frontend Engineer                & 4 yrs      & AI-era      & Feature Implementation    \\
P7  & F      & Mid-Level ML Engineer                      & 3 yrs      & AI-era      & Debugging                 \\
P8  & M      & Entry-Level R\&D Engineer                 & 1 yr       & AI-native   & Feature Implementation    \\
P9  & M      & Mid-Level R\&D Engineer                   & 2 yrs      & AI-era      & Debugging                 \\
P10 & F      & Mid-Level Software Engineer                & 2 yrs      & AI-era      & Feature Implementation    \\
P11 & M      & Full-Stack Developer                       & 1 yr       & AI-native   & Debugging                 \\
P12 & M      & Software Engineer                          & 3 yrs      & Pre-AI      & Debugging                 \\
P13 & M      & Entry-Level ML Engineer                    & 1 yr       & AI-native   & Declined\textsuperscript{\textdagger}                  \\
P14 & M      & Entry-Level Full-Stack Developer        & 1 yr       & AI-native   & Project Planning          \\
P15 & M      & Entry-Level Full-Stack Developer        & 1 yr       & AI-native   & Feature Implementation    \\ \hline
\multicolumn{6}{l}{\small{* \textit{The gender imbalance reflects broader trends in the software engineering profession and may affect the generalizability of findings.}}}\\
\multicolumn{6}{l}{\small{$\dagger$ \textit{P13 completed the interview but declined the coding task due to time constraints.}}}
\end{tabular}%
}
\end{table*}

\subsection{Interview Procedure}

Each interview followed a semi-structured guide designed to move from general reflections on security practices toward more specific questions about AI tool usage. We began by asking participants to describe how they approach security in their day-to-day work, whether they rely on any particular tools, frameworks, or mental habits, and how confident they feel about catching security issues before code ships. We then probed their use of AI coding assistants (including Claude Code, ChatGPT, Gemini, and Cursor) specifically, exploring whether and how these tools had changed their security workflows, any experiences where AI-generated code introduced or concealed a security problem, and how they evaluate the trustworthiness of AI output. Interviews were conducted remotely via video conferencing and recorded with explicit participant consent.

\subsection{Think-Aloud Coding Task}

Following the interview, participants were invited to complete a coding task using Gemini CLI with Gemini 2.5 Pro. Rather than assigning a fixed task, we offered 3 options aligned with common development activities: implementing a new feature (Task 1), initializing a new project (Task 2), or debugging existing code (Task 3). Participants selected whichever best matched their regular work to preserve ecological validity. One participant (P13) declined this component, and one participant (P5) attempted all of them.

To observe participant behavior without introducing the overhead of screen sharing software, we hosted a collaborative development environment using a self-hosted VS Code server~\cite{coder2026codeserver}. Participants accessed this environment through a shared link, and interviewers observed the session live from their own screen, recording their view throughout. Gemini CLI with Gemini 2.5 Pro was pre-installed and configured within the environment so that participants could begin working immediately without setup friction. This setup allowed us to capture prompting sequences, code edits, iteration patterns, and any security-relevant decisions or omissions in real time.

\subsubsection{Task Design Rationale}
The 3 coding tasks were designed to probe security awareness across distinct phases of the software development lifecycle. Each was grounded in a realistic scenario with no explicit security requirements, so that participants' natural reasoning would be observable rather than primed. Full task prompts are provided in Appendix~\ref{appendix:tasks}.

The \textbf{feature implementation} task asked participants to build a \texttt{/update-profile} HTTP endpoint accepting a JSON payload with \texttt{user\_id}, \texttt{display\_name}, and \texttt{bio}. It was designed to surface whether participants spontaneously applied input validation, authorization checks, or sanitization when handling untrusted user data, a persistent source of web application vulnerabilities~\cite{owasp2010} that developers frequently overlook even when aware of its importance~\cite{votipka2018hackers}

The \textbf{project initialization} task asked participants to scaffold a backend project with a configuration system, basic routing, and a \texttt{/health} endpoint. It probed whether participants spontaneously considered secure defaults, secret management, or environment separation at project inception, decisions that are substantially harder to retrofit later~\cite{mcgraw2012software}.

The \textbf{debugging} task presented a concurrency vulnerability in a Python file upload function that wrote all uploads to a shared temporary path before renaming, causing files to mix between users under simultaneous load. It probed reactive security reasoning: whether participants would recognize a latent race condition~\cite{bishop1996checking} as a security issue rather than treating it as a purely functional bug.

\subsection{Data Analysis}

\subsubsection{Interview Analysis}
Interview transcripts were analyzed using reflexive thematic analysis~\cite{braun2006using}. The first author conducted open coding of all transcripts, generating an initial set of codes grounded in participants' own language. The complete codebook, including all 20 codes across 6 thematic categories with definitions and representative quotes, is provided in Appendix~\ref{appendix:codebook}. These codes were then organized into candidate themes through iterative discussion with the broader research team. Analytical memos were maintained throughout to document interpretive decisions and track how themes evolved across the dataset. We paid particular attention to tensions between participants' stated security practices and the behaviors they described when using AI tools.

\subsubsection{Coding Session Analysis}
Screen recordings of the coding sessions were reviewed alongside the source code files produced during each session. Analysis focused on three dimensions: (1) prompting strategies and iteration patterns, including how participants refined or escalated prompts in response to AI output; (2) security-relevant decisions, including whether participants reviewed, questioned, or accepted AI-generated code that contained or omitted security considerations; and (3) task completion approach, including how participants structured their workflow and how much independent verification they performed. Coding session observations were triangulated with interview responses to identify participants whose stated practices aligned or diverged from their observed behavior.

\subsection{Limitations}
We want to acknowledge a few limitations in this study. The sample (N=15) is small and male-skewed, so the findings should be treated as exploratory rather than generalizable. The interview-first design may have primed participants to reflect on security before the coding task, though most still did not address security during it, suggesting any priming effect was minimal. We used Gemini CLI to ensure session consistency, which may not fully reflect other AI-native tools. Finally, the think-aloud setting may have introduced some deliberateness not present in routine work.

\section{Ethics Statement}
This study received approval from our Institution's Review Board (IRB) prior to data collection. All 15 participants were industry professionals recruited through purposive and snowball sampling via professional networks. Participation was voluntary and uncompensated. Participants provided informed consent, were told they could withdraw at any time, and are referred to by anonymous codes throughout. No identifiable personal data beyond voluntary demographics was collected.

\section{Findings}

We discovered a consistent and recurring tension across all participant groups: software engineers simultaneously rely on AI coding assistants as indispensable productivity tools while harboring persistent doubts about the security and reliability of the code they produce. Participants with more years of experience tended to articulate more specific security concerns and name vulnerability classes more precisely, while junior engineers more often framed security as something to catch in review rather than something to specify upfront. What united all groups, however, was an acknowledgment that AI has fundamentally changed the nature of software development work, even if its implications for code security remain personally and professionally unsettled.

\subsection{Productivity Gains and Shifting Development Workflows}

The most universally reported change brought about by AI coding assistants was a dramatic reduction in the time required to complete development tasks. Participants across all experience levels described AI as having compressed timelines that previously spanned days or weeks into hours. P2 captured this shift most directly, noting that tasks which once took a month or more could now be completed within a week. This acceleration was particularly pronounced for boilerplate code, utility functions, and routine feature implementation, tasks that participants consistently described as well-suited to AI delegation. Before the widespread availability of AI tools, participants described relying on a patchwork of resources, including Stack Overflow, documentation, YouTube tutorials, and online blogs, a process that AI has largely replaced with a single conversational interface. P5 noted that he still turns to Google first for simple problems but increasingly uses AI tools when a task requires planning, scoping, or exploration of unfamiliar libraries, explaining: \textit{``when I see it's a bit complicated, or when I see that there is some sort of planning required, I use it to figure out the scope and then work on it manually.''}

\subsubsection{The Junior Engineer Mental Model}

A particularly consistent pattern across participants was the framing of AI as a junior colleague rather than an authoritative system. P1 articulated this most explicitly, describing his approach as treating the AI \textit{``as my colleague who is slightly a junior engineer ...  I trust him, but not 100\%, I have to do my work and ensure credibility and reliability.''} P2 echoed this framing, stating that engineers should \textit{``treat AI output as a reference, not the final code, and as an assistant, not a replacement.''} This mental model appeared to function as a practical heuristic for managing trust: participants who held this view were more likely to describe careful review practices and less likely to report accepting suggestions wholesale. Notably, however, participants who endorsed this framing in the interview did not consistently translate it into security-oriented prompting behavior during the coding sessions, suggesting the mental model shapes how engineers describe their practice more reliably than it shapes the practice itself.

\subsubsection{Task Allocation and the Limits of AI Delegation}

The junior engineer framing also helps explain variation in how participants allocated AI assistance across task types, a pattern that became especially visible during the coding sessions. P3 described using AI primarily to generate logical outlines that he then rewrites almost entirely by hand, and during the session, he deliberately chose the project planning task, explaining that it was \textit{``the most fitting task for AI ...  it initializes backend, generates boilerplate, makes project runnable. This is something I would use AI for without any thought.''} He was equally explicit about what he would not delegate, stating that \textit{``the third task is something I would NOT use AI for ...  debugging something, I do not trust AI to do.''} P15 similarly avoided the debugging task on specification grounds, noting that \textit{``I don't want to debug something that isn't completely spec'd out.''} The consensus across this group was that AI performs best on narrowly scoped, well-defined tasks and degrades significantly when given broader autonomy over complex or underspecified problems.

\subsubsection{Changing Practices Around Code Review}

The introduction of AI into development workflows also altered how participants approached code review. P1 described a consistent practice of reading through AI-generated code before running it, checking the workflow, function calls, and parameter handling before proceeding to local testing and then deployment, noting that \textit{``if the AI gives me code, I first go through the code ...  what are the things happening, how is the workflow going through. If it seems reliable, then I run it locally.''} P6 emphasized testing as the primary verification mechanism, stating that \textit{``testing is the main thing for AI stuff ...  if you don't validate the code it wrote and go through the code changes, then it's not done. Manual testing is far better than AI testing.''} Several participants acknowledged that review practices tend to be more rigorous for production-bound code and more lenient for internal or experimental work, a distinction that introduces inconsistency into how thoroughly AI output is scrutinized across different development contexts.

\subsection{Trust Calibration and Security Awareness}

While productivity gains were celebrated, participants expressed pervasive concern about whether AI-generated code could be trusted from a security standpoint. The majority reported that they do not fully trust AI for security-sensitive implementation, and several described having caught specific vulnerabilities in AI-generated suggestions. P2 noted that AI tools tend to prioritize functionality and readability over security by default, observing that \textit{``the biggest concern for me is that code can look correct but still be insecure ...  AI often focuses on functionality and readability, not security by default.''} Trust ratings across participants clustered in a moderate range, with most assigning scores between 4 and 7 out of 10, reflecting a widespread sense that AI output requires verification but is not without value. Crucially, the coding sessions revealed a sharper picture: none of the 14 participants mentioned security in their initial prompts, and only 2 caught unstated security issues unprompted, a divergence between stated concern and observed behavior that held across experience levels.

\subsubsection{Vulnerability Detection in Practice}

The coding sessions surfaced meaningful variation in how participants detected, or failed to detect, security issues in AI-generated code. P10 identified a critical access control flaw in an AI-generated profile update endpoint immediately and without prompting. The AI had retrieved the current user ID from the POST body rather than a secure session store, meaning a malicious user could trivially impersonate any account by manipulating the request. P10 explained: \textit{``AI didn't implement any security feature ...  there is no check for that, so any user can update the profile. We need to implement that portion as AI didn't implement it.''} P12 took a different but equally effective approach, explicitly asking the AI whether there were additional security concerns after completing the initial fix. This prompted the AI to surface DDoS vulnerabilities that P12 had not considered: \textit{``it also handles some DDoS attack issues as well, which, to be honest, I didn't think of.''} These 2 participants were the only ones to catch unstated security issues across the 14 coding sessions. The remaining 12 participants either did not review for security or expressed satisfaction with outputs that contained unaddressed vulnerabilities, underscoring how substantially security outcomes varied based on individual prompting habits rather than on any consistent behavior the tool itself elicited.

\subsubsection{Experience Cohort and Security Performance}

A key motivating question for our study design was whether a developer's relationship to AI tools during professional formation would predict their security behavior in AI-assisted work. Our coding sessions did not support this expectation. The 2 participants who caught unstated security issues (P10 and P12) represent different cohorts (AI-era and Pre-AI, respectively) and different levels of seniority. Participants who missed critical security issues were similarly distributed across cohorts. This null result is informative: it suggests that experience level and AI familiarity, as we operationalized them, are not reliable proxies for security-aware AI use. What appeared to differentiate effective from ineffective security behavior was not years of experience or cohort membership but the specific dimension of security knowledge and whether participants had developed habits of security-oriented prompting. We return to the implications of this finding in the Discussion section.

\subsubsection{Role of Prompt Specificity in Security Outcomes}

A secondary theme that emerged from both interviews and coding sessions was the critical importance of prompt specificity in determining the security quality of AI output. P5 described constructing detailed prompt templates that included application structure, API patterns, and testing frameworks as a persistent context layer before issuing any task-specific instructions. P11 took a structured approach during the coding session, creating an agent configuration file that established code style rules and error handling expectations before engaging with the task itself, noting that this helped constrain the AI to a predictable and reviewable mode of operation. P10 observed that \textit{``the more context you give, the more refined the answer ...  but in the same chat, AI keeps the previous context and gives a mixture of both, even if you tell it not to,''} suggesting that prompt discipline must be maintained not only at the outset but throughout an entire session. The coding sessions reinforced this finding: participants who iterated on their prompts with security in mind produced more secure outcomes than those who issued a single functional prompt and accepted the result, though the sample is too small to characterize the magnitude of this effect.

\subsubsection{Patterns and Repetition in AI Suggestions}

Participants also noted that AI tools exhibit recognizable patterns in their suggestions that can create a false sense of familiarity and reliability. P2 observed that AI tools sometimes appear to retrieve cached responses, repeatedly offering the same solution regardless of how the prompt was varied. P10 described a related issue in which the AI confidently asserted implementation details that did not exist in the actual codebase: \textit{``sometimes it says garbage very confidently ...  it says that's how I can do it, but when I look into the codebase, I see there is no such thing like that.''} This pattern of confident but incorrect output was raised unprompted by multiple participants as one of the more practically dangerous properties of current tools, because the surface presentation of confidence provides no reliable signal of underlying accuracy.

\subsection{Diverging Views on Security Awareness Over Time}

One of the most contested questions across the interviews was whether AI tools make engineers more or less security-aware over time. Participants were genuinely divided, and several acknowledged the tension without resolving it. P3 was among the most candid, stating that AI tools represent a significant hindrance to his security awareness: \textit{``I become lazy with AI tools ...  security considerations take a backseat in my head. I would overlook things, not read line by line.''} He qualified this by noting that the productivity benefits still outweigh the security costs in his personal assessment, but the trade-off was acknowledged rather than dismissed. P5 extended this concern beyond his own practice, observing that \textit{``people are becoming less aware because they're not reviewing the code ...  they're just blindly accepting the code, which shouldn't be the case.''}

\subsubsection{The Risk of Uncritical Acceptance in Production Environments}

Several participants identified an emerging pattern of speed-driven code acceptance, where engineers accept AI output wholesale without meaningful review. P5 attributed this partly to organizational pressure, arguing that executives are pushing AI adoption without adequate understanding of its security limitations, warning that \textit{``many executives are really pushing it too much without understanding the problems and vulnerabilities ...  people are vibe coding production code, and that's really creating more risks.''} The coding sessions provided some direct evidence of this dynamic: 12 of the 14 participants included no security considerations in their initial prompts, and several expressed satisfaction with outputs that contained unaddressed vulnerabilities. P6 acknowledged that while he was satisfied with the output in the research context, \textit{``in actual job life, I wouldn't be''} satisfied with the same result, making explicit the gap between research task behavior and his own production standards. Whether this gap reflects the low-stakes research setting, time pressure, or something more persistent about how AI interaction is structured is a question our data cannot fully resolve.

\subsubsection{Concerns About Data Exposure and Sensitive Information}

A specific and recurring security concern across participants was the risk of inadvertently exposing sensitive information, either by sharing proprietary code or credentials with AI tools or by accepting AI-generated code that fails to handle sensitive data appropriately. P5 described an organizational context in which senior leadership initially discouraged external AI tools out of concern that proprietary information would be leaked, noting that \textit{``the models aren't really transparent about how they will use it, and malicious attackers might use that to steal information.''} P10 offered a practical framework for managing this risk, advising that engineers should share with AI only what does not expose client details, security keys, or architectural specifics, and concluding that \textit{``security vulnerabilities come from people, not from the tools ...  you need to be careful what you feed the AI.''} P11 independently raised hard-coded API keys and plain-text password storage as the clearest markers of insecure code in his own review process, noting that these are precisely the kinds of issues that are easy to miss when reviewing AI-generated output quickly.

\subsubsection{Education and Security Knowledge as Mediating Factors}

Participants consistently pointed to the engineer's own security knowledge as the primary determinant of whether AI-generated code ends up being secure or not. P4 described staying current on common vulnerability classes through blogs and capture-the-flag exercises, noting that this background knowledge is what allows him to recognize when AI output introduces broken access control, SQL injection risks, or supply chain vulnerabilities. The coding sessions reinforced this in a way that cuts against intuitive assumptions about experience: P10, despite having approximately 1.5 years of professional experience, caught a critical authorization flaw that more senior participants missed. Meanwhile, P12 (a more experienced Pre-AI developer) used a systematic post-implementation query strategy to surface additional vulnerabilities. Neither pattern aligns cleanly with cohort or seniority as predictors. Instead, both cases point toward specific security knowledge and deliberate verification habits as the operative factors. As P10 put it, \textit{``a person should cross-check it themselves because they know the way it should be implemented ...  AI doesn't.''} This convergence across participants suggests a troubling asymmetry: those most capable of catching AI-introduced vulnerabilities are those who already possess strong security knowledge, while those who most rely on AI may be least positioned to evaluate its security outputs critically.

\subsection{Attitudes Toward AI Autonomy and Agentic Tools}

As participants described their use of agentic AI tools, a distinct set of concerns emerged around the question of how much control engineers should delegate to systems capable of taking actions, modifying files, and executing commands without explicit step-by-step authorization. P3 described a moment during the coding session when he realized the AI agent could execute bash scripts on his local machine, which he found genuinely alarming: \textit{``it can execute a bash script on my PC, it can do anything it wants ...  that is a fairly strong security consideration for me.''} He stated that any indication of unsolicited scripting would cause him to halt the agent immediately and inspect what had occurred before proceeding, reflecting a principled commitment to maintaining human authority over system-level actions.

\subsubsection{Restricting Agent Permissions as a Coping Strategy}

Several participants described actively restricting the permissions available to their AI agents as a practical response to concerns about unintended actions. P9 described removing file write permissions from his agent after repeated experiences of the tool making changes he had not intended, framing this as preserving his own authority over the codebase: \textit{``I don't like to give you the autonomy to do what you want to do ...  maybe you're not right always.''} P11 took a more structured approach during the coding session, creating an agent configuration file before beginning the task, using it to define code style expectations and constrain the agent's behavior, noting that this gave him more confidence in reviewing the output that followed. P3 similarly described a preference for reviewing each proposed change individually rather than accepting batch edits, explaining that \textit{``I want nothing decided by it ...  I want full control over the code it generates.''} These strategies represent an emerging set of informal norms around agentic AI use that participants had developed independently, without formal guidance from their organizations.

\subsubsection{Redesigning AI Tools for Greater Security Integration}

When asked how they would redesign AI coding tools to better serve security goals, participants converged on several common themes. P2 argued for restricting the access AI tools have to project files and sensitive environment variables, reasoning that limiting exposure reduces the surface area for potential leakage. P5 emphasized efficiency and local execution, arguing that high API costs and cloud dependency create unnecessary risk and that tools should be redesigned to run on local hardware with modest specifications. P11 identified a more fundamental behavioral issue, observing that his preferred agentic tool would accept any input as ground truth regardless of its accuracy, and contrasting this unfavorably with tools that push back on incorrect premises: \textit{``I will 100\% change it to be more critical of the user's opinion.''} P12's practice during the coding session, explicitly asking the AI to identify remaining security concerns after completing a task, surfaced vulnerabilities that neither the engineer nor the initial AI output had raised, pointing to post-implementation security querying as a candidate design primitive. Across participants, the shared vision for improved AI tooling was one in which tools actively participate in quality and security enforcement rather than deferring entirely to engineer prompts, while preserving enough human oversight to catch errors before they reach production.

\section{Discussion}
Our findings reveal that the relationship between AI coding assistants and developer security awareness is neither straightforwardly harmful nor straightforwardly beneficial, but is instead deeply mediated by individual knowledge, prompting practice, and organizational context. This complexity resists simple characterization and demands a more nuanced account than either optimistic narratives of AI-assisted productivity or pessimistic warnings about automation complacency have so far provided.

\subsection{Security Awareness as a Function of the Individual, Not the Tool}

The most striking pattern in our coding sessions was how little the tool itself shaped security outcomes. None of the 14 participants included security requirements in their initial prompts, yet 2 caught critical vulnerabilities through independent knowledge and deliberate querying. P10, among the least experienced session participants, identified a fundamental authorization flaw that more senior developers missed. P12 surfaced additional vulnerabilities by explicitly asking the AI whether security concerns remained, a strategy no other participant employed. What differentiated these 2 participants was not seniority or AI familiarity but specific security knowledge applied at the right moment.

This suggests that AI coding assistants amplify existing security awareness rather than supplanting it. Engineers with strong security intuitions extend that intuition into their AI interactions; those without it accept functionally plausible output with no tool-level signal that review is warranted. This dynamic mirrors the asymmetry Perry et al.~\cite{perry2023users} observed, in which developers using AI produced less secure code while rating it more highly, and extends it: the gap is not just in outputs but in the capacity to evaluate them. As AI lowers the barrier to producing functional code, it may simultaneously raise the effective barrier to producing secure code, concentrating risk among the engineers who most depend on the tool. Given that current tools provide no mechanism to distinguish security-sensitive from routine generation~\cite{roch2024navigating}, this concentration is a structural outcome rather than an individual failure.

\subsection{The Prompting Gap and Its Consequences}

A related finding is what we term the prompting gap: the consistent absence of security requirements from initial AI interactions, even among engineers who articulated clear security concerns in the interview. This was not a knowledge failure. Participants who named common vulnerability classes and described authorization best practices did not spontaneously import that knowledge into their prompts. Security was treated as a subsequent concern to catch in review rather than a first-class requirement to specify upfront.

The structural explanation is that AI coding assistants are designed and evaluated around functional correctness. The prompt-then-suggest interaction model implicitly frames the engineer's role as functional specification; non-functional properties like security, accessibility, and performance fall outside that frame unless actively imported~\cite{anderson2010security}. Serafini et al.~\cite{serafini2025exploring} demonstrated experimentally that even lightweight structural interventions such as security prompts and warning messages meaningfully reduced insecure code use, which suggests the prompting gap is addressable at the interaction level. The implication is that fixing it requires reconsidering the interaction model itself, not simply encouraging engineers to write more thorough prompts.

\subsection{Informal Coping Strategies and Their Limits}

Without formal guidance or tool-level support, participants independently developed strategies for managing AI security risk: restricting agent permissions, creating configuration files to constrain agent behavior, reviewing changes individually rather than in bulk, and explicitly querying the AI for remaining security concerns after completing functional implementation. Where applied, these strategies were effective. Where absent, as was most often the case among less experienced participants, no mechanism existed to fill the gap.

The prevalence of these self-invented strategies points to a genuine practitioner need that current tools and organizations are not meeting. This pattern is consistent with Widder et al.'s~\cite{widder2023power} finding that software engineers' attempts to address quality concerns are constrained by organizational incentives and unsupported by institutional structures, leaving the burden on individual initiative. The informal strategies our participants developed, and particularly the practice of post-implementation security querying, are candidate patterns for formalization. Evaluated and incorporated into both tool design and training curricula, they could make security-aware AI use consistent rather than contingent on individual resourcefulness.

\subsection{Toward Security-Aware AI Coding Assistants}

Our findings point toward design directions that address structural conditions rather than layering security warnings onto existing interaction patterns. Three directions warrant particular attention.

First, security should be a first-class concern in the generation workflow. This could mean automatic security review passes following code generation, prompts requiring engineers to specify security requirements before generation begins in sensitive contexts, or inline flagging of patterns associated with OWASP vulnerability classes~\cite{owasp2021}. The goal is to make security consideration structurally embedded rather than optional, shifting the default from functional to holistic specification.

Second, trust calibration mechanisms should be built into the review interface. Our participants applied similar confidence levels to boilerplate code and security-sensitive logic alike, with no tool signaling that the stakes differed. Lee et al.~\cite{lee2024one} found that exposing users to multiple potentially inconsistent outputs reduced overreliance by making model uncertainty visible. Applying this principle to security-critical code sections by surfacing alternative implementations, flagged assumptions, or explicit confidence signals would help engineers calibrate scrutiny to context rather than applying uniform trust across all generated code~\cite{wang2024investigating}.

Third, organizations must treat security-aware AI use as a competency requiring deliberate investment. Several participants described workplace cultures in which adoption pressure outpaced the development of security review practices, and in which the effort required for careful AI use was structurally discouraged. Madaio et al.~\cite{madaio2024learning} found that practitioners need sociotechnical understanding of AI rather than narrowly technical guidance, which suggests that training should address the behavioral and organizational dimensions of AI-assisted development alongside the technical ones. Shared prompt templates, review checklists encoding the informal strategies our participants developed, and explicit policies governing what information may be shared with AI tools would institutionalize practices that currently exist only in individual workflows. Tool design alone cannot create these conditions, but guidance framing security-aware AI use as an organizational competency rather than a property that tools deliver automatically is a tractable first step.

\section{Design Implications}

Our findings, taken together with the structural gaps identified in the existing literature, point toward concrete design directions for AI coding assistants, developer tooling, and organizational practice. We organize these implications around three levels at which intervention is both warranted and feasible.

\subsection{Tools Should Treat Security as a Core Interaction Concern}

The most consequential design change would be to restructure the AI coding assistant interaction model so that security is surfaced proactively rather than left entirely to the engineer. Current tools present code generation as a functional task, implicitly framing security as a post-hoc concern. This framing is not neutral: it shapes behavior, as our coding sessions demonstrated, with zero participants including security requirements in initial prompts despite being able to articulate security concerns when directly asked. Serafini et al. \cite{serafini2025exploring} demonstrated experimentally that security prompts and warning messages reduce the use of insecure AI-generated code, suggesting that even lightweight structural interventions at the point of interaction can shift behavior meaningfully. Tools should implement automatic security review passes following code generation, require engineers to confirm security requirements before generation begins for sensitive code contexts, and surface inline warnings for patterns commonly associated with vulnerability classes, such as those documented by OWASP \cite{owasp2021}. The goal is not to add friction indiscriminately but to make security considerations a structural part of the generation workflow rather than an optional addition.

\subsection{Trust Calibration Mechanisms Should Be Built Into the Review Interface}

Poorly calibrated trust, rather than trust itself, poses the core design challenge our findings surface. When developers apply uniform confidence across boilerplate and security-sensitive logic alike, the interface bears partial responsibility: presenting a single best suggestion signals equivalence where none exists. AI coding assistants should differentiate output presentation by security sensitivity, surfacing alternative implementations, and flagging embedded assumptions precisely where the cost of misplaced trust is highest.

This reframing shifts the intervention from the developer to the tool. Wang et al.~\cite{wang2024investigating} found that developers ground trust in perceived competence and integrity, yet current interfaces offer little to validate either. Making the AI's security reasoning visible at the point of review would let engineers evaluate not just functional correctness but the soundness of the assumptions underneath it, turning code review from passive acceptance into active scrutiny.

\subsection{Organizations Should Formalize Security-Aware AI Use as a Competency}

Tool-level interventions alone are insufficient if the organizational conditions surrounding AI adoption continue to reward speed over scrutiny. Several participants described workplace cultures where pressure to ship code quickly discouraged the additional effort required for security aware AI use. In these environments, informal coping strategies such as permission restriction, iterative security querying, and agent configuration were developed individually rather than shared or institutionalized. Earlier research \cite{widder2023power} found that engineers’ attempts to address quality and ethical concerns were constrained by organizational incentives and often psychologically taxing, indicating that individual vigilance is not a sustainable substitute for institutional support. Organizations adopting AI coding assistants should treat security-aware prompting as a formal competency. They should develop shared prompt templates and review checklists that capture the informal strategies our participants described. They should also establish clear policies about what information may be shared with AI tools. Research shows \cite{madaio2024learning} that practitioners need a sociotechnical understanding of AI rather than narrowly technical guidance. Therefore, training should address behavioral and organizational dimensions of AI-assisted development in addition to technical skills. Taken together, these three levels of intervention reflect a recognition that the security awareness gap we identified is not a property of individual engineers but a structural consequence of how AI coding assistants are designed, deployed, and governed.

\section{Conclusion}
This paper set out to examine not whether AI coding assistants generate vulnerable code, but how they reshape the security awareness of the developers who use them. The answer is nuanced. AI does not erode security knowledge, but it reorganizes when that knowledge is applied, shifting it from a preventive stance embedded in writing code to a reactive one exercised during review. That shift is a predictable consequence of interaction models that treat code generation as a functional task and leave security to be addressed afterward. Three findings stand out: security knowledge and prompting behavior are decoupled across experience cohorts; informal coping strategies point to a practitioner need that neither tools nor organizations are meeting; and experience level shows no reliable relationship to security performance during coding sessions. The goal is not to slow adoption but to shape it. Tool designers need to treat security as primary in the generation workflow, organizations need to formalize security-aware AI use as a competency rather than assuming it emerges naturally, and researchers need to keep examining how these tools transform practice in ways vulnerability benchmarks cannot capture.

\section{Acknowledgment}
This work was supported in part by NSF grants CCF-2348277 and CCF-2518445 \cite{zaman2024crii}.

\bibliographystyle{plain}
\bibliography{usenix2024_SOUPS}

\appendix

\section{Codebook for Interview Analysis}
\label{appendix:codebook}

Table~\ref{tab:codebook} presents the full codebook developed through open and focused coding of the 15 interview transcripts. Codes are organized into 6 thematic categories that emerged inductively from the data, yielding 20 codes in total. Each entry includes the code name, a definition, and a representative example quote from a participant.

\begin{table*}[t]
\small
\centering
\caption{Interview Codebook: Categories, Codes, Definitions, and Example Quotes}
\label{tab:codebook}
\begin{tabular}{p{3cm} p{2.5cm} p{4.0cm} p{6.0cm}}
\hline
\textbf{Category} & \textbf{Code} & \textbf{Definition} & \textbf{Example Quote} \\ 
\hline

Security Practices and Habits
& Preventive security behavior
& Proactive security measures integrated during development (e.g., validation, sanitization, access control).
& \textit{``I always validate what comes in from the client before I do anything with it.''} (P1) \\ \hline

Security Practices and Habits
& Reactive security behavior
& Security actions taken only after issues are detected via testing or reports.
& \textit{``Usually I find out there's a problem when someone files a ticket.''} (P12) \\ \hline

Security Practices and Habits
& Security tooling
& Use of automated tools (e.g., static analysis, dependency scanning) in the workflow.
& \textit{``We run Snyk in CI, so vulnerabilities get caught before anything ships.''} (P3) \\ \hline

Security Practices and Habits
& Security knowledge gaps
& Self-reported uncertainty about vulnerability types or mitigation practices.
& \textit{``I know SQL injection is a thing, but I wouldn't know how to test for it.''} (P8) \\ \hline

AI Tool Usage and Trust
& Uncritical AI acceptance
& Integration of AI-generated code with minimal review for correctness or security.
& \textit{``If it runs and tests pass, I usually just commit it.''} (P15) \\ \hline

AI Tool Usage and Trust
& Critical AI evaluation
& Active inspection or modification of AI output before integration.
& \textit{``I always read what it gives me. I've caught things that looked fine.''} (P1) \\ \hline

AI Tool Usage and Trust
& AI-induced complacency
& Reduced vigilance attributed to reliance on AI suggestions.
& \textit{``It should know about these things better than me.''} (P11) \\ \hline

Security Awareness Shift with AI
& Increased security awareness
& Reported growth in security attentiveness due to AI feedback.
& \textit{``Sometimes it suggests something I hadn't thought of.''} (P6) \\ \hline

Security Awareness Shift with AI
& Decreased security awareness
& Reported decline in deliberate security reasoning after AI adoption.
& \textit{``I used to think through edge cases more.''} (P9) \\ \hline

Security Awareness Shift with AI
& No perceived change
& Security habits reported as unchanged despite AI use.
& \textit{``I do the same things I always did.''} (P4) \\ \hline

Prompting and Interaction Behavior
& Security-explicit prompting
& Explicit inclusion of security requirements in AI prompts.
& \textit{``Make sure this handles invalid input.''} (P5) \\ \hline

Prompting and Interaction Behavior
& Security-absent prompting
& Prompts focused only on functionality without security constraints.
& \textit{``I just describe what I want it to do.''} (P14) \\ \hline

Prompting and Interaction Behavior
& Iterative security correction
& Follow-up prompts to correct security issues in AI output.
& \textit{``I told it to add an auth check.''} (P10) \\ \hline

Experience and Cohort Effects
& Pre-AI security grounding
& Security mental models formed prior to AI tool adoption.
& \textit{``I learned this stuff before these tools existed.''} (P3) \\ \hline

Experience and Cohort Effects
& AI-era hybrid approach
& Combined reliance on prior training and AI assistance.
& \textit{``I know enough to know when it's wrong.''} (P7) \\ \hline

Experience and Cohort Effects
& AI-native dependency
& Heavy reliance on AI with limited independent security reasoning.
& \textit{``I haven't really worked without it.''} (P8) \\ \hline

Perceived Responsibility and Ownership
& Personal security ownership
& Explicit acknowledgment of individual accountability for code security.
& \textit{``If something is insecure, that's on me.''} (P5) \\ \hline

Perceived Responsibility and Ownership
& Diffused responsibility
& Ambiguity about accountability when AI-generated code is involved.
& \textit{``I'm not sure who's responsible.''} (P11) \\ \hline

Perceived Responsibility and Ownership
& Organizational security reliance
& Dependence on team processes rather than individual initiative.
& \textit{``That's what code review is for.''} (P9) \\ \hline

Perceived Responsibility and Ownership
& Security as a secondary concern
& Security framed as subordinate to functionality or deadlines.
& \textit{``If the feature isn't working, nothing else matters.''} (P12) \\ \hline

\end{tabular}
\end{table*}

\clearpage

\section{Interview Protocol}
\label{appendix:protocol}
The following semi-structured interview guide was used across all sessions.
Questions were adapted conversationally based on participant responses.
Interviews averaged 35 minutes.

\subsection*{Part A: Background and Experience}
\begin{itemize}
\item How long have you been developing software professionally?
\item What types of applications or projects do you typically work on?
\item When did you first start using AI coding assistants, and which tools have you used?
\item How frequently do you use AI coding assistants in your development work?
\item Have you received any formal training in secure coding practices or application security?
\item How confident do you feel about catching security issues before code ships?
\end{itemize}

\subsection*{Part B: AI Tool Usage and Trust}
\begin{itemize}
\item Walk me through your typical workflow when using an AI coding assistant.
\item How do you decide when to use AI assistance versus coding something manually?
\item How much do you trust code generated by AI tools, and what influences that level of trust?
\item How do you evaluate whether AI-generated code is correct and safe to use?
\item Can you recall a time when an AI tool suggested code that was problematic or insecure? How did you recognize the issue?
\item Do you review AI-generated code differently than code you write yourself?
\end{itemize}

\subsection*{Part C: Security Practices and AI}
\begin{itemize}
\item How has using AI coding assistants changed how you think about security, if at all?
\item Can you describe a situation where AI-generated code introduced or concealed a security problem?
\item Are there security checks or practices you apply consistently regardless of whether code is AI-generated?
\item Do you feel more or less confident about the security of code when AI was involved in writing it?
\end{itemize}

\subsection*{Part D: Coding Task Reflection}
\begin{itemize}
\item Walk me through the security considerations you had in mind during the task.
\item Did the AI assistant help or complicate security implementation in this task?
\item Are there security features you knew you should implement but did not? Why not?
\item If you were to do this task again, what would you do differently from a security standpoint?
\end{itemize}

\subsection*{Closing}
\begin{itemize}
\item Is there anything about AI coding tools and security that we did not discuss but you think is important?
\end{itemize}

\section{Coding Task Problem Statements}
\label{appendix:tasks}

The following task descriptions were provided to participants.
No security-specific requirements were included. Participants were free
to use any language or framework.

\subsection*{Task 1: Feature Implementation}

\textbf{Prompt.}
Create an HTTP endpoint called \texttt{/update-profile} that allows a user
to update their profile information. The client will send JSON containing:

{\ttfamily
\begin{flushleft}
\{
  "user\_id": 123, \\
  "display\_name": "Example Name", \\
  "bio": "Example Bio" \\
\}
\end{flushleft}
}

\textbf{Requirements.}
\begin{itemize}
\item Implement the endpoint in the language and framework of your choice.
\item Update the user's profile in a persistent store or mock database.
\item Handle the fields \texttt{user\_id}, \texttt{display\_name}, and \texttt{bio}.
\item Return a JSON response indicating success or failure.
\item Include any validation or request handling logic you consider appropriate.
\end{itemize}

\subsection*{Task 2: Project Initialization}

\textbf{Prompt.}
Plan and initialize a backend project in the language and framework of your
choice. The project should run with simple commands and include a basic
structure for configuration and routing.

\textbf{Requirements.}
\begin{itemize}
\item Create a project folder with all necessary setup files.
\item Add a configuration system for managing environment-specific settings.
\item Implement a simple \texttt{/health} endpoint returning a status message.
\item Include instructions for running the project and supplying the configuration.
\item Provide a minimal but extensible project layout.
\end{itemize}

\subsection*{Task 3: Debugging}

\textbf{Prompt.}
Users report that when multiple people upload files simultaneously,
files occasionally end up in the wrong user's folder. The following
function is used to save uploaded files:

{\ttfamily
\begin{flushleft}
def save\_file(user\_id, uploaded\_file): \\
\quad temp\_path = "/tmp/upload.tmp" \\
\quad with open(temp\_path, "wb") as f: \\
\qquad f.write(uploaded\_file.read()) \\
\quad final\_path = f"/data/\{user\_id\}/\{uploaded\_file.filename\}" \\
\quad os.rename(temp\_path, final\_path)
\end{flushleft}
}

\textbf{Requirements.}
\begin{itemize}
\item Identify why files may be mixed between users.
\item Modify the function so that simultaneous uploads do not interfere.
\item Ensure files are saved under the correct user's directory.
\item Ensure the revised version works reliably under concurrent uploads.
\end{itemize}

\end{document}